\documentclass[12pt]{article}
\usepackage{amssymb,amsmath,epsfig}
\begin{document}
\title{\bf Some Bianchi Type Cosmological Models in $f(R)$ Gravity}

\author{M. Farasat Shamir \thanks{frasat@hotmail.com}\\\\
Department of Mathematics, University of the Punjab,\\
Quaid-e-Azam Campus, Lahore-54590, Pakistan.}

\date{}

\maketitle
\begin{abstract}
The modified theories of gravity, especially the $f(R)$ gravity,
have attracted much attention in the last decade. In this context,
we study the exact vacuum solutions of Bianchi type $I,~III$ and
Kantowski-Sachs spacetimes in the metric version of $f(R)$
gravity. The field equations are solved by taking expansion scalar
$\theta$ proportional to shear scalar $\sigma$ which gives
$A=B^n$, where $A$ and $B$ are the metric coefficients. The
physical behavior of the solutions has been discussed using some
physical quantities. Also, the function of the Ricci scalar is
evaluated in each case.
\end{abstract}


\section{Introduction}

The late time accelerated expansion of the universe has attracted
much attention in the recent years. Direct evidence of cosmic
acceleration comes from high red-shift supernova experiments
\cite{1}. Some other observations, such as cosmic microwave
background fluctuations \cite{2} and large scale structure
\cite{3}, provide an indirect evidence. These observations seem to
change the entire picture of our matter filled universe. It is now
believed that most part of the universe contains dark matter and
dark energy. The modifications of general relativity (GR) seem
attractive to explain late time acceleration and dark energy.

Among the various modifications of GR, the $f(R)$ theory of
gravity is treated most seriously during the last decade. It
provides a natural gravitational alternative to dark energy. It
has been suggested that cosmic acceleration can be achieved by
replacing the Einstein-Hilbert action of GR with a general
function Ricci scalar $f(R)$. $f(R)$ theory of gravity  has been
shown \cite{01} equivalent to scalar-tensor theory of gravity that
is incompatible with solar system tests of GR, as long as the
scalar field propagates over solar system scales. The explanation
of cosmic acceleration is obtained just by introducing the term
$1/R$ which is essential at small curvatures. Capozziello et al.
\cite{02} have shown that dust matter and dark energy phases can
be achieved by the exact solution derived from a power law $f(R)$
cosmological model. The $f(R)$ theory of gravity is considered
most suitable due to cosmologically important $f(R)$ models. These
models consist of higher order curvature invariants as functions
of the Ricci scalar. Viable $f(R)$ gravity models \cite{4} have
been proposed which show the unification of early-time inflation
and late-time acceleration. The problem of dark matter can also be
addressed by using viable $f(R)$ gravity models. There are some
other useful aspects \cite{5} of $f(R)$ gravity. It gives an easy
unification of early time inflation and late time acceleration. It
can be used for the explanation of hierarchy problem in high
energy physics. It also describes the transition phase of the
universe from deceleration to acceleration. Thus $f(R)$ theory of
gravity seems attractive and a reasonable amount of work has been
done in different contexts.

Lobo and Oliveira \cite{06} constructed wormhole geometries in the
context of $f(R)$ theories of gravity. Cognola et al. \cite{6}
investigated $f(R)$ gravity at one-loop level in de-Sitter
universe. It was found that one-loop effective action can be
useful for the study of constant curvature black hole nucleation
rate. Multam$\ddot{a}$ki and Vilja \cite{7} investigated
spherically symmetric vacuum solutions in $f(R)$ theory. The same
authors \cite{8} also studied the perfect fluid solutions and
showed that pressure and density did not uniquely determine
$f(R)$. In a recent paper, Sharif and Kausar \cite{09} studied
non-vacuum static spherically symmetric solutions in this theory.
Capozziello et al. \cite{9} explored spherically symmetric
solutions of $f(R)$ theories of gravity via the Noether symmetry
approach. Hollenstein and Lobo \cite{10} analyzed exact solutions
of static spherically symmetric spacetimes in $f(R)$ gravity
coupled to non-linear electrodynamics. Azadi et al. \cite{11}
studied cylindrically symmetric vacuum solutions in this theory.
Momeni and Gholizade \cite{12} extended cylindrically symmetric
solutions in a more general way. Reboucas and Santos \cite{013}
studied G$\ddot{o}$del-type universes in f(R) gravity. We have
explored static plane symmetric vacuum solutions \cite{13} in
$f(R)$ gravity. The field equations are solved using the
assumption of constant scalar curvature which may be zero or
non-zero. In a recent paper, Babichev and Langlois \cite{130}
studied relativistic stars in this theory.

Friedmann-Robertson-Walker (FRW) models, being spatially
homogeneous and isotropic in nature, are best for the
representation of the large scale structure of the present
universe. However, it is believed that the early universe may not
have been exactly uniform. Thus, the models with anisotropic
background are the most suitable to describe the early stages of
the universe. Bianchi type models are among the simplest models
with anisotropic background. Many authors \cite{14}-\cite{19}
explored Bianchi type spacetimes in different contexts. Kumar and
Singh \cite{019} studied solutions of the field equations in the
presence of perfect fluid using Bianchi type $I$ spacetime in GR.
Moussiaux et al. \cite{20} investigated the exact solution for
vacuum Bianchi type-$III$ model with a cosmological constant.
Lorenz-Petzold \cite{14} studied exact Bianchi type-$III$
solutions in the presence of electromagnetic field. Xing-Xiang
\cite{21} discussed Bianchi type $III$ string cosmology with bulk
viscosity. He assumed that the expansion scalar is proportional to
the shear scalar to derive the solutions. Wang \cite{022}
investigated string cosmological models with bulk viscosity in
Kantowski-Sachs spacetime. Upadhaya \cite{22} explored some
magnetized Bianchi type-$III$ massive string cosmological models
in GR. Recently, Hellaby \cite{023} presented an overview of some
recent developments in inhomogeneous models and it was concluded
that the universe is inhomogeneous on many scales.

The investigation of Bianchi type models in alternative or
modified theories of gravity is also an interesting discussion.
Kumar and Singh \cite{23} investigated perfect fluid solutions
using Bianchi type $I$ spacetime in scalar-tensor theory. Singh et
al. \cite{24} studied some Bianchi type-$III$ cosmological models
in scalar-tensor theory. Adhav et al. \cite{25} obtained an exact
solution of the vacuum Brans-Dicke field equations for the metric
tensor of a spatially homogeneous and anisotropic model. Paul et
al. \cite{26} investigated FRW cosmologies in $f(R)$ gravity.
Recently, we \cite{27,28} have studied the solutions of Bianchi
types $I$ and $V$ spacetimes in the framework of $f(R)$ gravity.

In this paper, we focuss our attention to explore the vacuum
solutions of Bianchi types $I,~III$ and Kantowski-Sachs spacetimes
in metric $f(R)$ gravity. The field equations are solved by taking
expansion scalar $\theta$ proportional to shear scalar $\sigma$
which gives $A=B^n$, where $A$ and $B$ are the metric
coefficients. The paper is organized as follows: A brief
introduction of the field equations in metric version of $f(R)$
gravity is given in section \textbf{2}. In section \textbf{3},
\textbf{4} and \textbf{5}, the solutions of the field equations
for Bianchi types $I,~III$ and Kantowski-Sachs spacetimes are
found. Some physical parameters and the functions of Ricci scalar
are also evaluated in the context of these solutions. In the last
section, we discuss the results.

\section{Some Basics of $f(R)$ Gravity}

The metric tensor plays an important role in GR. The dependence of
Levi-Civita connection on the metric tensor is one of the main
properties of GR. However, if we allow torsion in the theory, then
the connection no longer remains the Levi-Civita connection and
the dependence of connection on the metric tensor vanishes. This
is the main idea behind different approaches of $f(R)$ theories of
gravity.

When the connection is the Levi-Civita connection, we get metric
$f(R)$ gravity. In this approach, we take variation of the action
with respect to the metric tensor only. The action for $f(R)$
gravity is given by
\begin{equation}\label{1}
S=\int\sqrt{-g}(f(R)+L_{m})d^4x,
\end{equation}
where $f(R)$ is a general function of the Ricci scalar and $L_{m}$
is the matter Lagrangian. The field equations resulting from this
action are
\begin{equation}\label{2}
F(R)R_{\mu\nu}-\frac{1}{2}f(R)g_{\mu\nu}-\nabla_{\mu}
\nabla_{\nu}F(R)+g_{\mu\nu}\Box F(R)=T^m_{\mu\nu},
\end{equation}
where $F(R)\equiv
df(R)/dR,~\Box\equiv\nabla^{\mu}\nabla_{\mu},~\nabla_{\mu}$ is the
covariant derivative and $T^m_{\mu\nu}$ is the standard minimally
coupled stress energy tensor derived from the Lagrangian $L_m$.
Now contracting the field equations, it follows that
\begin{eqnarray*}
F(R)R-2f(R)+3\Box F(R)=T
\end{eqnarray*}
In vacuum, this reduces to
\begin{equation}\label{3}
F(R)R-2f(R)+3\Box F(R)=0.
\end{equation}
which implies that
\begin{equation}\label{4}
f(R)=\frac{3\Box F(R)+F(R)R}{2}.
\end{equation}
This gives an important relationship between $f(R)$ and $F(R)$
which will be used to simplify the field equations and to evaluate
$f(R)$.

Now we consider the metric
\begin{equation}\label{5}
ds^{2}=dt^2-A^2(t)dr^2-B^2(t)[d{\theta}^2+l^2(\theta)d{\phi}^2],
\end{equation}
where
\[ l^2(\theta) = \left\lbrace
  \begin{array}{c l}
    {\theta}^2 & \text{~~~~~when~~k=0~~~(Bianchi I model),}\\
    \sinh^2{\theta} & \text{~~~~~when~~k=-1~~(Bianchi III
model),}\\\sin^2{\theta} & \text{~~~~~when~~k=1~~~(Kantowski-Sachs
model).}
  \end{array}
\right. \]

Here $k$ is the spatial curvature index and the above three models
are Euclidian, semi-closed and closed respectively.

\section{Bianchi Type $I$ Solution}

Here we shall find exact solutions of Bianchi $I$ spacetime in
$f(R)$ gravity. For the sake of simplicity, we take the vacuum
field equations. The line element of Bianchi type $I$ spacetime is
given by
\begin{equation}\label{6}
ds^{2}=dt^2-A^2(t)dr^2-B^2(t)[d{\theta}^2+\theta^2d\phi^2],
\end{equation}
where $A$ and $B$ are cosmic scale factors. The corresponding
Ricci scalar is given by
\begin{equation}\label{7}
R=-2[\frac{\ddot{A}}{A}+\frac{2\ddot{B}}{B}+
\frac{2\dot{A}\dot{B}}{AB}+\frac{\dot{B}^2}{B^2}],
\end{equation}
where dot denotes derivative with respect to $t$. Using
Eq.(\ref{4}), the vacuum field equations take the form,
\begin{equation}\label{8}
\frac{F(R)R_{\mu\nu}-\nabla_{\mu}\nabla_{\nu}F(R)}{g_{\mu\nu}}
=\frac{F(R)R-\Box F(R)}{4}.
\end{equation}
One can view Eq.(\ref{8}) as the set of differential equations for
$F(t)$, $A$ and $B$. It follows from Eq.(\ref{8}) that the
combination
\begin{equation}\label{9}
A_{\mu}\equiv\frac{F(R)R_{\mu\mu}-\nabla_{\mu}\nabla_{\mu}
F(R)}{g_{\mu\mu}},
\end{equation}
is independent of the index $\mu$ and hence $A_{\mu}-A_{\nu}=0$
for all $\mu$ and $\nu$. Thus $A_{0}-A_{1}=0$
gives
\begin{equation}\label{10}
-\frac{2\ddot{B}}{B}+\frac{2\dot{A}\dot{B}}{AB}
+\frac{\dot{A}\dot{F}}{AF}-\frac{\ddot{F}}{F}=0.
\end{equation}
Also, $A_{0}-A_{2}=0$ yields
\begin{equation}\label{11}
-\frac{\ddot{A}}{A}-\frac{\ddot{B}}{B}
+\frac{\dot{A}\dot{B}}{AB}+\frac{\dot{B}^2}{B^2}
+\frac{\dot{B}\dot{F}}{BF}-\frac{\ddot{F}}{F}=0.
\end{equation}
Now we give definition of some physical quantities before solving
these equations.

The average scale factor $a$ and the volume scale factor $V$ are
defined as
\begin{equation}\label{12}
a=\sqrt[3]{AB^2},\quad V=a^3=AB^2.
\end{equation}
The average Hubble parameter $H$ is given in the form
\begin{equation}\label{13}
H=\frac{1}{3}(\frac{\dot{A}}{A}+\frac{2\dot{B}}{B}).
\end{equation}

The expansion scalar $\theta$ and shear scalar $\sigma$ are
defined as follows
\begin{eqnarray}\label{14}
\theta&=&u^\mu_{;\mu}=\frac{\dot{A}}{A}+2\frac{\dot{B}}{B},\\
\label{15} \sigma^2&=&\frac{1}{2}\sigma_{\mu\nu}\sigma^{\mu\nu}
=\frac{1}{3}[\frac{\dot{A}}{A}-\frac{\dot{B}}{B}]^2,
\end{eqnarray}
where
\begin{equation}\label{16}
\sigma_{\mu\nu}=\frac{1}{2}(u_{\mu;\alpha}h^\alpha_\nu+u_{\nu;\alpha}h^\alpha_\mu)
-\frac{1}{3}\theta h_{\mu\nu},
\end{equation}
$h_{\mu\nu}=g_{\mu\nu}-u_{\mu}u_{\nu}$ is the projection tensor
while $u_\mu=\sqrt{g_{00}}(1,0,0,0)$ is the four-velocity in
co-moving coordinates.

We have two differential equations given by
Eqs.(\ref{10},\ref{11}) with three unknowns namely $A,~B$ and $F$.
Thus we need one additional constraint to solve these equations.
We use a physical condition that expansion scalar $\theta$ is
proportional to shear scalar $\sigma$ which gives
\begin{equation}\label{17}
A=B^n.
\end{equation}
Using this condition, Eqs.(\ref{10},\ref{11}) take the form
\begin{eqnarray} \label{18}
-\frac{2\ddot{B}}{B}+2n\frac{\dot{B}^2}{B^2}
+n\frac{\dot{B}\dot{F}}{BF}-\frac{\ddot{F}}{F}&=&0,\\\label{19}
(n+1)\frac{\ddot{B}}{B}+(n^2-2n-1)\frac{\dot{B}^2}{B^2}-
\frac{\dot{B}\dot{F}}{BF}+\frac{\ddot{F}}{F}&=&0.
\end{eqnarray}
Adding these, we get
\begin{equation}\label{20}
\frac{\ddot{B}}{B}+(n+1)\frac{\dot{B}^2}{B^2}
+\frac{\dot{B}\dot{F}}{BF}=0.
\end{equation}
We solve this equation using power law relation between $F$ and
$a$ \cite{27},
\begin{equation}\label{21}
F=ka^m,
\end{equation}
where $k$ is the constant of proportionality, $m$ is any
integer and $a$ is given by
\begin{equation}\label{22}
a=B^{\frac{n+2}{3}}
\end{equation}
Thus for $m=3$, we obtain
\begin{equation}\label{23}
F=kB^{n+2}.
\end{equation}
Using this in Eq.(\ref{20}), it follows that
\begin{equation}\label{24}
\frac{\ddot{B}}{B}+(2n+3)\frac{\dot{B}^2}{B^2}=0.
\end{equation}
Put $\dot{B}=g(B)$ in this equation, we get
\begin{equation}\label{25}
\frac{dg^2}{dB}+\frac{4n+6}{B}g^2=0,
\end{equation}
which leads to the solution
\begin{equation}\label{26}
g^2=\frac{c_1}{B^{4n+6}},
\end{equation}
where $c_1$ is an integration constant. Hence the solution becomes
\begin{equation}\label{27}
ds^{2}=(\frac{dt}{dB})^2dB^2-B^{2n}dr^2-B^2(d\theta^2+\theta^2d\phi^2),
\end{equation}
which can be written as
\begin{equation}\label{28}
ds^{2}=\frac{1}{c_1}B^{4n+6}dB^2-B^{2n}dr^2-B^2(d\theta^2+\theta^2d\phi^2),
\end{equation}
and after the transformations $B=T,~r=R,~\theta=\Theta$ and
$\phi=\Phi$, it takes the form
\begin{equation}\label{29}
ds^{2}=\frac{1}{c_1}T^{4n+6}dT^2-T^{2n}dR^2-T^2(d\Theta^2+\Theta^2d\Phi^2).
\end{equation}

The average Hubble parameter becomes
\begin{equation}\label{30}
H=\frac{\sqrt{c_1}(n+2)}{3T^{2n+4}}.
\end{equation}
while the volume scale factor turns out to be
\begin{equation}\label{31}
V=T^{n+2}.
\end{equation}
The expansion scalar $\theta$ is given by
\begin{equation}\label{32}
\theta=\frac{\sqrt{c_1}(n+2)}{T^{2n+4}},
\end{equation}
while the shear scalar $\sigma$ becomes
\begin{equation}\label{33}
\sigma^2=\frac{c_1(n-1)^2}{3T^{4n+8}}.
\end{equation}

Moreover, the function of Ricci scalar, $f(R)$, can be found by
using Eq.(\ref{4})
\begin{equation}\label{34}
f(R)=\frac{kT^{n+2}R}{2}.
\end{equation}
It follows from Eq.(\ref{7}) that
\begin{equation}\label{35}
R=2c_1(n^2+6n+5)T^{-4n-8}.
\end{equation}
Thus $f(R)$ can be written as a function of $R$ only
\begin{equation}\label{36}
f(R)=\frac{k}{2}[2c_1(n^2+6n+5)]^{\frac{n+2}{4n+8}}R^{\frac{3}{4}}.
\end{equation}

\section{Bianchi Type $III$ Solution}

Here we shall find the solution of the Bianchi type $III$
spacetime in $f(R)$ gravity for the vacuum field equations. The
line element of Bianchi type $III$ spacetime is given by
\begin{equation}\label{37}
ds^{2}=dt^2-A^2(t)dr^2-B^2(t)[d{\theta}^2+\sinh^2\theta d\phi^2],
\end{equation}
where $A$ and $B$ are cosmic scale factors. The Ricci scalar for
this spacetime is given by
\begin{equation}\label{38}
R=-2[\frac{\ddot{A}}{A}+\frac{2\ddot{B}}{B}+
\frac{2\dot{A}\dot{B}}{AB}-\frac{1}{B^2}+\frac{\dot{B}^2}{B^2}],
\end{equation}
where dot denotes derivative with respect to $t$. Using
Eq.(\ref{9}), the vacuum field equations take the form,
\begin{eqnarray}\label{39}
-\frac{2\ddot{B}}{B}+\frac{2\dot{A}\dot{B}}{AB}
+\frac{\dot{A}\dot{F}}{AF}-\frac{\ddot{F}}{F}=0,\\\label{40}
-\frac{\ddot{A}}{A}-\frac{\ddot{B}}{B}
+\frac{\dot{A}\dot{B}}{AB}-\frac{1}{B^2}+\frac{\dot{B}^2}{B^2}
+\frac{\dot{B}\dot{F}}{BF}-\frac{\ddot{F}}{F}=0.
\end{eqnarray}
Here we also need one additional constraint to solve these
equations. Using the same physical condition that expansion scalar
$\theta$ is proportional to shear scalar $\sigma$, we obtain
\begin{eqnarray} \label{41}
-\frac{2\ddot{B}}{B}+2n\frac{\dot{B}^2}{B^2}
+n\frac{\dot{B}\dot{F}}{BF}-\frac{\ddot{F}}{F}&=&0,\\\label{42}
(n+1)\frac{\ddot{B}}{B}+(n^2-2n-1)\frac{\dot{B}^2}{B^2}-
\frac{\dot{B}\dot{F}}{BF}+\frac{\ddot{F}}{F}&=&-\frac{1}{B^2}.
\end{eqnarray}
Adding these, we get
\begin{equation}\label{43}
\frac{\ddot{B}}{B}+(n+1)\frac{\dot{B}^2}{B^2}
+\frac{\dot{B}\dot{F}}{BF}=-\frac{1}{(n-1)B^2}.
\end{equation}
Using power law relation between $F$ and $a$, we obtain
\begin{equation}\label{44}
F=kB^{n+2}.
\end{equation}
Thus we get a differential equation with one unknown,
\begin{equation}\label{45}
\frac{\ddot{B}}{B}+(2n+3)\frac{\dot{B}^2}{B^2}=-\frac{1}{(n-1)B^2}.
\end{equation}
Put $\dot{B}=g(B)$ in this equation, we get
\begin{equation}\label{46}
\frac{dg^2}{dB}+\frac{4n+6}{B}g^2=-\frac{2}{(n-1)B},
\end{equation}
which leads to the solution
\begin{equation}\label{47}
g^2=\frac{c_2}{B^{4n+6}}-\frac{1}{(n-1)(2n+3)},
\end{equation}
where $c_2$ is an integration constant. Hence the solution becomes
\begin{equation}\label{48}
ds^{2}=(\frac{1}{\frac{c_2}{B^{4n+6}}-\frac{1}{(n-1)(2n+3)}})dB^2-
B^{2n}dr^2-B^2(d\theta^2+\sinh^2\theta \phi^2),
\end{equation}
which takes the form
\begin{equation}\label{49}
ds^{2}=(\frac{1}{\frac{c_2}{T^{4n+6}}-\frac{1}{(n-1)(2n+3)}})dT^2-
T^{2n}dR^2-T^2(d\Theta^2+\sinh^2\theta \phi^2).
\end{equation}
where $B=T,~r=R,~\theta=\Theta$ and $\phi=\Phi$.

The average Hubble parameter becomes here
\begin{equation}\label{50}
H=(\frac{n+2}{3})[\frac{c_2}{T^{4n+8}}-\frac{1}{(n-1)(2n+3)T^2}]^{\frac{1}{2}}.
\end{equation}
while the volume scale factor turns out to be same as for the
Bianchi type $I$ spacetime. The expansion scalar $\theta$ is given
by
\begin{equation}\label{51}
\theta=(n+2)[\frac{c_2}{T^{4n+8}}-\frac{1}{(n-1)(2n+3)T^2}]^{\frac{1}{2}},
\end{equation}
while the shear scalar $\sigma$ becomes
\begin{equation}\label{52}
\sigma^2=\frac{1}{3}(n-1)^2[\frac{c_2}{T^{4n+8}}-\frac{1}{(n-1)(2n+3)T^2}].
\end{equation}

Moreover, the function of Ricci scalar, $f(R)$, can be found by
using Eq.(\ref{4})
\begin{equation}\label{53}
f(R)=\frac{k}{2}[T^{n+2}R-\frac{3(2n^2+7n+6)}{2n^2+n-3}T^n].
\end{equation}
It follows from Eq.(\ref{38}) that
\begin{equation}\label{54}
R=2[\frac{c_2(n^2+6n+5)}{T^{4n+8}}+\frac{1}{T^2}(\frac{3n^2+2n-2}{2n^2+n-3})],
\end{equation}
which clearly indicates that $f(R)$ cannot be explicitly written
in terms of $R$. However, for a special case when $n=-5$, $f(R)$
turns out to be
\begin{equation}\nonumber
f(R)=\frac{k}{12\sqrt{3}}R^{5/2}
\end{equation}
This gives $f(R)$ only as a function of $R$.

\section{Kantowski-Sachs Solution}

The line element of Kantowski-Sachs spacetime is
\begin{equation}\label{55}
ds^{2}=dt^2-A^2(t)dr^2-B^2(t)[d{\theta}^2-\sin^2\theta d\phi^2],
\end{equation}
where $A$ and $B$ are cosmic scale factors. The corresponding
Ricci scalar is given by
\begin{equation}\label{56}
R=-2[\frac{\ddot{A}}{A}+\frac{2\ddot{B}}{B}+
\frac{2\dot{A}\dot{B}}{AB}+\frac{1}{B^2}+\frac{\dot{B}^2}{B^2}],
\end{equation}
where dot denotes derivative with respect to $t$. The vacuum field
equations for Kantowski-Sachs spacetime are given by
\begin{eqnarray}\label{57}
-\frac{2\ddot{B}}{B}+\frac{2\dot{A}\dot{B}}{AB}
+\frac{\dot{A}\dot{F}}{AF}-\frac{\ddot{F}}{F}=0,\\\label{58}
-\frac{\ddot{A}}{A}-\frac{\ddot{B}}{B}
+\frac{\dot{A}\dot{B}}{AB}+\frac{1}{B^2}+\frac{\dot{B}^2}{B^2}
+\frac{\dot{B}\dot{F}}{BF}-\frac{\ddot{F}}{F}=0.
\end{eqnarray}
Here we also use Eq.(\ref{17}) so that the field equations take
the form
\begin{eqnarray} \label{59}
-\frac{2\ddot{B}}{B}+2n\frac{\dot{B}^2}{B^2}
+n\frac{\dot{B}\dot{F}}{BF}-\frac{\ddot{F}}{F}&=&0,\\\label{60}
(n+1)\frac{\ddot{B}}{B}+(n^2-2n-1)\frac{\dot{B}^2}{B^2}-
\frac{\dot{B}\dot{F}}{BF}+\frac{\ddot{F}}{F}&=&\frac{1}{B^2}.
\end{eqnarray}
Adding these, we get
\begin{equation}\label{61}
\frac{\ddot{B}}{B}+(n+1)\frac{\dot{B}^2}{B^2}
+\frac{\dot{B}\dot{F}}{BF}=\frac{1}{(n-1)B^2}.
\end{equation}
Using Eq.(\ref{23}), we obtain one differential equation with one
unknown,
\begin{equation}\label{62}
\frac{\ddot{B}}{B}+(2n+3)\frac{\dot{B}^2}{B^2}=\frac{1}{(n-1)B^2},
\end{equation}
which gives
\begin{equation}\label{63}
{\dot{B}}^2=\frac{c_3}{B^{4n+6}}+\frac{1}{(n-1)(2n+3)},
\end{equation}
where $c_3$ is an integration constant. Hence the solution becomes
\begin{equation}\label{64}
ds^{2}=(\frac{1}{\frac{c_3}{T^{4n+6}}+\frac{1}{(n-1)(2n+3)}})dT^2-
T^{2n}dR^2-T^2(d\Theta^2+\sin^2\theta \phi^2).
\end{equation}
where $B=T,~r=R,~\theta=\Theta$ and $\phi=\Phi$.

The average Hubble parameter becomes here
\begin{equation}\label{65}
H=(\frac{n+2}{3})[\frac{c_3}{T^{4n+8}}+\frac{1}{(n-1)(2n+3)T^2}]^{\frac{1}{2}}.
\end{equation}
while the volume scale factor turns out to be same as for the
Bianchi types $I$ and $III$  spacetimes. The expansion scalar
$\theta$ is given by
\begin{equation}\label{66}
\theta=(n+2)[\frac{c_3}{T^{4n+8}}+\frac{1}{(n-1)(2n+3)T^2}]^{\frac{1}{2}},
\end{equation}
while the shear scalar $\sigma$ turns out to be
\begin{equation}\label{67}
\sigma^2=\frac{1}{3}(n-1)^2[\frac{c_3}{T^{4n+8}}+\frac{1}{(n-1)(2n+3)T^2}].
\end{equation}

The function of Ricci scalar becomes
\begin{equation}\label{68}
f(R)=\frac{k}{2}[T^{n+2}R+\frac{3(2n^2+7n+6)}{2n^2+n-3}T^n],
\end{equation}
while Ricci scalar takes the form
\begin{equation}\label{69}
R=2[\frac{c_3(n^2+6n+5)}{T^{4n+8}}-\frac{1}{T^2}(\frac{3n^2+2n-2}{2n^2+n-3})].
\end{equation}
For a special case when $n=-5$, $f(R)$ turns out to be
\begin{equation}\nonumber
f(R)=\frac{k}{4\sqrt{3}}|-R|^{5/2}.
\end{equation}

\section{Concluding Remarks}

The main purpose of this paper is to study some Bianchi type
cosmological models in metric $f(R)$ gravity. We find exact
solution of the vacuum field equations for Bianchi type $I,~III$
and Kantowski-Sachs spacetimes. Initially, the field equations
look complicated but lead to a solution using some assumptions.
The first assumption is that the expansion scalar $\theta$ is
proportional to the shear scalar $\sigma$. It gives $A=B^n$, where
$A,~B$ are the metric coefficients and $n$ is an arbitrary
constant. Secondly, the power law relation between $F$ and $a$ is
used to find the solution. Some important cosmological physical
quantities for the solutions such as expansion scalar $\theta$,
shear scalar $\sigma^2$ and average Hubble parameter are
evaluated. The analysis of these parameters in the light of solar
system constraints is not done in this paper. However, it would be
worthwhile to check the consistency of these parameters with
Wilkinson Microwave Anisotropy Probe (WMAP) data. Also, the
solutions can be compared with Lemaître-Tolman cosmology
\cite{023} that describes the inhomogeneity in the universe on
many scales.

The general function of Ricci scalar, $f(R)$, is also constructed
in each case. For some special cases, it is found that $f(R)$
models include some root powers of the Ricci scalar. It has been
shown \cite{29} that the model $f(R)=R^{3/2}$ is consistent with
cosmological results. In particular, it is possible to obtain flat
rotation curves for galaxies and consistency with solar system
tests. However, more work has still to be done to find the
viability of these models and consistency with solar system.

The models of the universe in Eqs.(\ref{29},\ref{49},\ref{64}) are
non-singular at $T=0$. The physical parameters $H,~\theta$ and
$\sigma$ are all infinite at this point but the volume scale
factor $V$ vanishes. The general function of the Ricci scalar is
finite while the spatial part of the metric vanish at $T=0$. The
expansion stops for $n=-2$ in all models. The models indicate that
after a large time the expansion will stop completely and the
universe will achieve isotropy. The isotropy condition, i.e.,
$\frac{\sigma^2}{\theta}\rightarrow 0$ as $T\rightarrow \infty$,
is also satisfied in each case. Thus we can conclude from these
observations that the models start their expansion from zero
volume at $T=0$ and the volume increases with the passage of time.

\end{document}